\newcommand{\avg}[1]{\left< #1 \right>} 
\newcommand{\fe}{\mathrm{Fe^{11+}}}
\documentclass[manuscript]{aastex}
\usepackage{amsbsy}
\usepackage{amsfonts}
\usepackage{amssymb}
\usepackage{longtable}
\usepackage{lscape}

\usepackage{graphicx}
\usepackage{graphics}
\usepackage{dcolumn}
\usepackage{bm}

\begin{document}


\title{ Storage Ring Cross Section Measurements for Electron Impact Ionization of $\mathrm{\mathbf{Fe^{11+}}}$ Forming $\mathrm{\mathbf{Fe^{12+}}}$ and $\mathrm{\mathbf{Fe^{13+}}}$}

\author{M. Hahn\altaffilmark{1}, D. Bernhardt\altaffilmark{2}, M. Grieser\altaffilmark{3}, C. Krantz\altaffilmark{3},  M. Lestinsky\altaffilmark{1}, A. M\"{u}ller\altaffilmark{2}, \\  O. Novotn\'{y}\altaffilmark{1}, R. Repnow\altaffilmark{3},
S. Schippers\altaffilmark{2}, A. Wolf\altaffilmark{3}, and D. W. Savin\altaffilmark{1}}

\altaffiltext{1}{Columbia Astrophysics Laboratory, Columbia University,\\ 550 West 120th, New York, NY 10027 USA}
\altaffiltext{2}{Institut f\"{u}r Atom- und Molek\"{u}lphysik, Justus-Liebig-Universit\"{a}t Giessen,\\ Leihgesterner Weg 217, 35392 Giessen, Germany}
\altaffiltext{3}{Max-Planck-Institut f\"{u}r Kernphysik,\\ Saupfercheckweg 1, 69117 Heidelberg, Germany}

\date{\today}
\begin{abstract}

	We report ionization cross section measurements for electron impact single ionization (EISI) of $\mathrm{Fe^{11+}}$ forming $\mathrm{Fe^{12+}}$ and electron impact double ionization (EIDI) of $\mathrm{Fe^{11+}}$ forming $\mathrm{Fe^{13+}}$. The measurements cover the center-of-mass energy range from approximately 230~eV to 2300~eV. The experiment was performed using the heavy ion storage ring TSR located at the Max-Planck-Institut f\"{u}r Kernphysik in Heidelberg, Germany. The storage ring approach allows nearly all metastable levels to relax to the ground state before data collection begins. We find that the cross section for single ionization is $30\%$ smaller than was previously measured in a single pass experiment using an ion beam with an unknown metastable fraction. We also find some significant differences between our experimental cross section for single ionization and recent distorted wave (DW) calculations. The DW Maxwellian EISI rate coefficient for $\fe$ forming $\mathrm{Fe^{12+}}$ may be underestimated by as much as 25\% at temperatures for which $\fe$ is abundant in collisional ionization equilibrium. This is likely due to the absence of $3s$ excitation-autoionization (EA) in the calculations. However, a precise measurement of the cross section due to this EA channel was not possible because this process is not distinguishable experimentally from electron impact excitation of an $n=3$ electron to levels of $n \geq 44$ followed by field ionization in the charge state analyzer after the interaction region. Our experimental results also indicate that the double ionization cross section is dominated by the indirect process in which direct single ionization of an inner shell $2l$ electron is followed by autoionization resulting in a net double ionization.

\end{abstract}
	
\maketitle
	
\section{Introduction}

	Collisionally ionized atomic plasmas are formed in a range of astrophysical objects including stellar coronae, galaxies, and supernova remnants. The charge state distribution (CSD) of such plasmas is determined by a balance between electron impact ionization (EII) and electron-ion recombination. The CSD plays an important role in a wide range of spectroscopic diagnostics used to infer electron temperature, electron density, and elemental abundances \citep{Brickhouse:AIP:1996, Landi:AA:1999, Bryans:ApJ:2009}. Most CSD calculations are carried out under the assumption of collisional ionization equilibrium (CIE) for conditions where the electron density is low (meaning three body recombination is unimportant), radiation can be ignored, and there is no dust. In this case the ionization and recombination rates are the same so that $n_{\mathrm{e}} n_{q} \alpha_{\mathrm{I}}^{q} = n_{\mathrm{e}} n_{q+1} \alpha_{\mathrm{R}}^{q+1}$, where $n_{\mathrm{e}}$ is the electron density, $n_{q}$ is the density of ions of a particular element with charge $q$, $\alpha_{\mathrm{I}}^{q}$ is the rate coefficient for ionization from $q$ to $q+1$, and $\alpha_{\mathrm{R}}^{q+1}$ is the rate coefficient for recombination from $q+1$ to $q$. Rewriting gives $n_{q+1}/n_{q} = \alpha_{\mathrm{I}}^{q} / \alpha_{\mathrm{R}}^{q+1}$, thus making clear the importance of accurate ionization and recombination data to model the CSD of a plasma.

	For CIE, generally only electron impact single ionization (EISI) need be considered since for a given charge state multiple ionization is significant only at temperatures so high that the fractional abundance of that ion is negligible \citep{Tendler:PhysLett:1984}. Multiple-electron ionization, such as electron impact double ionization (EIDI), need be considered only when modeling the CSD in dynamic systems where ions are suddenly exposed to higher electron temperatures \citep{Muller:PhysLett:1986}. Examples of such non-equilibrium systems include solar flares \citep{Reale:ApJ:2008}, supernova remnants \citep{Patnaude:ApJ:2009}, and merging galaxy clusters \citep{Akahori:PASJ:2010}. 
	
	Much work has been carried out in deriving the necessary EII data \citep[see the compilation by][]{Dere:AA:2007}, however sizable discrepencies in the data exist even among the most recent compilations \citep{Bryans:ApJ:2009}. Experimental EII measurements, like those reported here, will help to resolve these differences. These measurements can also be used to benchmark theory enabling more accurate EII cross section calculations for ions isoelectronic to those measured.
	
		The difficulty of producing well characterized ground state ion beams has been a major limitation for obtaining the accurate EII data needed for CIE models. Most measurements of EII have been performed in a single-pass geometry in which metastable ions generally have not had enough time to radiatively relax to the ground state before the EII measurements are performed. In this paper we describe an experiment employing an ion storage ring. This arrangement allows the ions to be stored long enough for typically all the metastable levels to radiatively relax before data acquisition begins. This fact has been exploited previously by \citet{Linkemann:PRL:1995, Linkemann:NIMB:1995}.
	
	Here we study ionization of $\fe$. This ion produces strong spectral lines which can be used for plasma diagnostics and instrument calibration \citep{DelZanna:AA:2005}. It is also a particularly important ion for observations of the solar corona \citep{Moses:SolPhys:1997, Brown:ApJS:2008}. 
	
	EISI of $\fe$ has already been the subject of some theoretical and experimental work \citep{Younger:JQSRT:1983, Pindzola:PRA:1986,Gregory:PRA:1987, Dere:AA:2007}. \citet{Gregory:PRA:1987} measured the ionization cross section for $\fe$ using a single pass experimental geometry. The EISI cross section measured by that experiment was about 30\% larger than that predicted by the distorted wave calculations of \citet{Pindzola:PRA:1986} or \citet{Dere:AA:2007}. \citet{Gregory:PRA:1987} compared their experimental result to theoretical calculations for ionization from the ground and metastable levels and concluded that the discrepancy was due to a large metastable population in the ion beam in the experiment. The measurements reported here, by being able to generate nearly pure ground state ion beams, can resolve this issue and also provide an excellent test case to compare crossed beams experimental results with the storage ring technique.

	We study in detail EISI of P-like $\fe$ forming Si-like $\mathrm{Fe^{12+}}$ over the electron-ion collision energy range of 230~$-$~2300~eV. This energy range includes the following direct ionization channels: 
\begin{equation}
\mathrm{e^{-} + Fe^{11+}} (2s^2\, 2p^6\, 3s^2\, 3p^3) \rightarrow \left\{ \begin{array}{l} 
													\mathrm{Fe^{12+}} (2s^2\, 2p^6\, 3s^2\, 3p^2) + 2\mathrm{e^{-}} \\
													\mathrm{Fe^{12+}} (2s^2\, 2p^6\, 3s\, 3p^3) + 2\mathrm{e^{-}} \\
													\mathrm{Fe^{12+}} (2s^2\, 2p^5\, 3s^2\, 3p^3) + 2\mathrm{e^{-}} \\
													\mathrm{Fe^{12+}} (2s^1\, 2p^6\, 3s^2\, 3p^3) + 2\mathrm{e^{-}} 
													\end{array} \right. .
\label{eq:transitions}
\end{equation}
The energy thresholds for direct ionization are 330.79~eV from the $3p$ subshell and 357.40~eV from the $3s$ subshell \citep{NIST:2008}. Excitation-autoionization (EA) can also occur starting at the ionization threshold of $330.79$~eV through the electron impact excitation of a $3s$ electron to an autoionizing state. We are unaware of any theoretical work on this specific EA channel for $\fe$. 
	
	Theoretical calculations have shown that EA is expected to make an important contribution to the ionization cross section above the excitation threshold for $n=2$ electrons at $\approx 710$~eV \citep{Pindzola:PRA:1986}. In that same energy range resonant processes such as resonant excitation double autoionization (REDA) and resonant excitation auto-double-ionization (READI) are predicted in the cross section. These occur through dielectronic capture when an ion forms an excited state that subsequently decays by ejecting two electrons \citep{Laguttuta:PRA:1981,Henry:PRA:1982,Muller:PRL:1988b,Muller:PRL:1988,Muller:Book:2008}. In the REDA process the electrons are released sequentially, whereas in READI the two electrons are ejected simultaneously. 

	The theoretical energy thresholds for ionization of the $2p$ or $2s$ electrons are 1073~eV and 1199~eV, respectively \citep{Kaastra:AAS:1993}. Ionization from the $2p$ subshell radiatively stabilizes with a predicted probability of $1.1 \%$ and from the $2s$ with a probability of $8.0\%$ \citep{Kaastra:AAS:1993}. In both cases it is more likely that direct ionization of an $n=2$ electron will be followed by autoionization giving a net double ionization to form $\mathrm{Fe^{13+}}$. Thus, ionization of an inner shell electron is expected to be a very small contribution to the total single ionization cross section, but it can be significant for double ionization \citep{Muller:PRL:1980}.

	Electron impact double ionization of $\fe$ forming Al-like $\mathrm{Fe^{13+}}$ was measured over the same energy range of 230 -- 2300~eV. The energy threshold for EIDI through direct ionization is 691.83~eV \citep{NIST:2008}. As discussed above, double ionization can also occur through single ionization of an inner shell electron with subsequent autoionization. Other indirect processes such as EA and resonant processes analogous to those expected for single ionization may also occur in the double ionization cross section \citep{Muller:PRL:1988}.

	The rest of this paper is organized as follows: In Section~\ref{sec:ex} we describe the experimental setup. The data analysis is presented in Section~\ref{sec:ana} and uncertainties in Section~\ref{sec:err}. Experimental results and comparison to theory are presented for single ionization in Section~\ref{sec:sres} and for double ionization in Section~\ref{sec:dres}. A summary is given in Section~\ref{sec:sum}.
	
\section{Experimental Setup}\label{sec:ex}

	 EII measurements were performed at the TSR heavy ion storage ring of the Max-Plank-Institut f\"{u}r Kernphysik in Heidelberg, Germany.  The experiments basically followed the procedure of previous experiments \citep{Kilgus:PRA:1992,Linkemann:PRL:1995, Kenntner:NIMPR:1995, Schippers:ApJ:2001, Lestinsky:ApJ:2009}. Details related specifically to ionization are described in \citet{Hahn:ApJ:2010}. Here we describe additional aspects relevant to the present work. 
	
	The TSR facility is equipped with two separate electron-ion merged beams sections. For the results presented here the electron beam commonly referred to as the Cooler \citep{Steck:NIMA:1990,Pastuszka:NIMA:1996} was used as a probe beam for electron-ion collision studies at tunable relative energies. The second electron beam device is known as the Target \citep{Sprenger:NIMA:2004}. This device was operated as an electron cooler with the energy fixed at what is referred to as the cooling energy \citep{Poth:PhysRep:1990}. Fixing the Target electron beam energy allowed the ion beam to be cooled continuously. This inhibits expansion of the beam due to warming and counteracts the drag force, both of which arise as the electron energy of the probe beam is varied during measurement.

	A beam of 150~MeV $\mathrm{^{56}Fe^{11+}}$ ions was injected into TSR in a series of 5 pulses spaced 0.8~s apart. The ion beam was merged with the two electron beams described above. Initially both the Cooler and Target electron beams were set to the space-charge-corrected cooling energy of 1460~eV. The center-of-mass energy spread is limited by the Cooler electron temperature, which can be described by a flattened Maxwellian distribution with temperatures in the perpendicular and parallel directions of typically $k_{\mathrm{B}}T_{\perp} = 13.5$~meV and $k_{\mathrm{B}}T_{\parallel}=180$~$\mu\mathrm{eV}$ (Kilgus et al. 1992; Schippers et al. 2001; Novotn\'y et al., in preparation). During measurement, the average ion current was 1 -- 2~$\mathrm{\mu A}$.
	
	A delay of 2 -- 3~s followed the last injection pulse before data collection began. This allowed $\fe$ metastable states to decay. To estimate the metastable fraction during measurement we modelled the level populations as a function of storage time starting from a Boltzmann distribution with temperature 750~eV, corresponding to the approximate collision energy of the electrons in the stripping foil as the ions pass through the foil in the accelerator. The energy levels and Einstein coefficients of \citet{NIST:2008} were used for the calculation. The longest lived metastable level was the $^{2}D_{5/2}$ level within the ground configuration, which has a lifetime of about $0.5$~s \citep{NIST:2008}.  The metastable fraction of the ion beam at the end of the initial cooling cycle is estimated to be $< 0.5\%$. 
	
	Products of ionizing electron-ion collisions formed in the Cooler were separated from the $\fe$ parent beam by the first dipole magnet downstream of the interaction section. Recombination and ionization were measured simultaneously using separate detectors on opposite sides of the parent ion beam. The detectors were positioned for maximum signal collection by stepping each unit horizontally and vertically in small increments across each product beam. Single or double ionization events were detected by positioning the ionization detector to intercept either the $\mathrm{Fe^{12+}}$ or $\mathrm{Fe^{13+}}$ product beam. 
	
	The ionization detector employs a channel electron multiplier \citep[CEM;][]{Rinn:RSI:1982, Linkemann:NIMB:1995}. A suitable CEM discriminator level was determined by measuring the total pulse height distribution. The level was set so that the detection efficiency of ionization events was essentially unity. The total ionization detector count rate was $\lesssim 20$~kHz with a typical pulse width of 30~ns. As a result, the signal loss induced by the dead time of the CEM was negligible. 
	
	Data acquisition began after the injection and initial cooling cycle already described. During data collection the relative energy between the probe electron beam and the ion beam was varied. Each energy scan consists of $\sim$~250 -- 700 pairs of steps, one step at the measurement energy and the other at a fixed reference energy. The total duration of each step was 12 -- 35~ms. There was a delay of 5 -- 20~ms at the beginning of each energy step before data were collected in order to allow the power supply to settle at the new voltage. Data were collected for the remaining 5 -- 15~ms of each step.
	
	The laboratory electron energies were always chosen to be higher than the cooling energy and fell between 4000~and~8000~eV.  Each energy scan covered a range of 80 -- 2000~V in the laboratory frame. A fast high-voltage amplifier with a dynamic range of $\pm 1000$~V was used to quickly switch voltages for the energy scans. This high-voltage amplifier was used in combination with a slower power supply to lift the fast amplifier into the voltage range desired for the energy scan, during which the slower power supply maintained a constant voltage. 
	
	The number of pairs of steps in the scan and the range of the energy scan were chosen to balance the desired energy resolution against the lifetime of the ion beam, which was about 18~s. The energy scans used fell into three broad categories. Low resolution overview scans, covering the entire 2000~V range permitted by the fast amplifier, revealed the general shape of the cross section. Medium resolution scans, covering an $\approx 330$~eV laboratory energy range, were used to capture major features of the cross section, such as EA, and to fill in the details in the overview scan. Finally, high resolution scans, covering a range of $\approx 80$~eV in the laboratory frame, were used to resolve even finer details such as resonances. The medium and high resolution scans were normalized to the low resolution scans to correct for ion current measurement offsets as described in Section~\ref{sec:err}. 
	
	In between every two measurement steps a reference step was used to estimate the background by measuring the count rate at a fixed reference energy. For single ionization measurements the reference energy was set below the single ionization threshold for energy scans up to a center-of-mass energy of about 950~eV. In this case the single ionization count rate at the reference energy was only due to single electron stripping (SES) off the residual gas. In the case of double ionization, energy scans extending to 1400~eV were performed with the reference below the double ionization threshold so that the count rate at reference was due only to double electron stripping (DES). At energies above 950 (1400)~eV the limited dynamic range of the fast high voltage amplifier prevented us from setting the reference point below the single (double) ionization threshold. For these higher energies the reference point was set to an energy where the cross section had already been determined from lower energy scans. For all data runs the background count rate due to stripping at the reference energy was corrected to better represent the background at the measurement energy using the method discussed in the next section. 
	
	Cycles of ion injection, cooling, and energy scan were repeated for $\sim 1$~hour to improve the statistical accuracy. After many cycles the process was repeated for a new energy range. Energy ranges were chosen to maintain at least 50\% overlap with other scans for high statistical accuracy and to correct for systematic offsets due to fluctuations in the ion current measurement calibration.

\section{Data Analysis}\label{sec:ana}

	The data analysis follows the procedure described in \citet{Hahn:ApJ:2010}. To review briefly, the cross section for single ionization $\sigma_{\mathrm{I}}$ versus energy is obtained from the measured ionization rate coefficients $\avg{\sigma_{\mathrm{I}}v_{\mathrm{rel}}}$, which are averaged over the velocity spread $\Delta v_{\mathrm{rel}}$ of the experiment. Because the center-of-mass energy spread is very small, $\Delta v_{\mathrm{rel}} \ll v_{\mathrm{rel}}$ and the cross sections can be calculated by dividing the averaged rate coefficients by the relative velocities. The cross section for single ionization, in terms of measured quantities, is therefore (see Appendix~\ref{ap:ref})
\begin{equation}
\sigma_{\mathrm{I}}(E_{\mathrm{m}}) = \frac{1}{v_{\mathrm{rel}}} \left\{\frac{ R_{\mathrm{I}}^{\mathrm{m}}(E_{\mathrm{m}}) - R_{\mathrm{I}}^{\mathrm{b}}(E_{\mathrm{m}})}{[1 - \beta_{\mathrm{i}}\beta{\mathrm{e}}(E_{\mathrm{m}})] n_{\mathrm{e}}^{\mathrm{m}}N_{\mathrm{i}}^{\mathrm{m}}L} + \avg{\sigma_{\mathrm{I}}v_{\mathrm{rel}}}(E_{\mathrm{r}})\frac{n_{\mathrm{e}}^{\mathrm{r}}}{n_{\mathrm{e}}^{\mathrm{m}}}
\frac{[1-\beta_{\mathrm{i}}\beta_{\mathrm{e}}(E_{\mathrm{r}})]}{[1-\beta_{\mathrm{i}}\beta_{\mathrm{e}}(E_{\mathrm{m}})]}
 \right\}.
\label{eq:sigmacalc}
\end{equation}	
Here $R_{\mathrm{I}}^{\mathrm{m}}(E_{\mathrm{m}})$ denotes the total single ionization count rate at the measurement step; $R_{\mathrm{I}}^{\mathrm{b}}(E_{\mathrm{m}})$ denotes the background single ionization count rate at the measurement step, which is proportional to the count rate at the reference step $R_{\mathrm{I}}^{\mathrm{r}}(E_{\mathrm{r}})$ as described in \citet{Hahn:ApJ:2010}; $L=1.5$~m is the length of the interaction region for the probe beam; the factor $(1-\beta_{\mathrm{i}}\beta{\mathrm{e}})$ is a relativistic correction where $\beta_{\mathrm{i}}$ and $\beta_{\mathrm{e}}$ are the ion and electron velocities normalized by the speed of light, $v_{\mathrm{i}}/c$ and $v_{\mathrm{e}}/c$, respectively; $E_{\mathrm{m}}$ and $E_{\mathrm{r}}$ are the center-of-mass energies at measurement and reference, respectively. The electron density $n_{\mathrm{e}}$ is of the order of $10^{7}$~$\mathrm{cm^{-3}}$ and is calculated from the measured electron current and the geometry of the probe beam \citep{Kilgus:PRA:1992}. 	The total number of stored ions per unit length $N_{\mathrm{i}}$ is calculated from the measured ion current. The EISI rate at the reference energy $\avg{\sigma_{\mathrm I}v_{\mathrm{rel}}}(E_{\mathrm{r}})$ is included in order to remove the EISI contribution to the background count rate $R_{\mathrm{I}}^{\mathrm{b}}$ when the reference energy is above the EISI threshold. The factor $n_{\rm e}^{\rm r}/n_{\rm e}^{\rm m}$ accounts for the different electron densities at reference and measurement. The EIDI cross section $\sigma_{\mathrm{DI}}$ can be calculated in the same manner as the EISI cross section by replacing the single ionization count rates in equation~(\ref{eq:sigmacalc}) with the analogous double ionization rates and the term $\avg{\sigma_{\mathrm I}v_{\mathrm{rel}}}(E_{\mathrm r})$ by the equivalent double ionization term $\avg{\sigma_{\mathrm{DI}}v_{\mathrm{rel}}}(E_{\mathrm r})$. 
	
	For all measurements, the electron and ion beams also interact in the merging and demerging sections on either side of the straight section in which the electron probe and ion beam co-propagate. In these sections the ion and electron beams meet at an angle causing the relative velocity to be greater than when the beams are colinear. These toroidal effects are accounted for using the method of \citet{Lampert:PRA:1996} to correct an overestimate of the cross section of $\approx$ 20\%. 
	
	The total ionization detector signal at the measurement energy is made up of EISI or EIDI plus a stripping background. We estimate the background $R_{\mathrm{I}}^{\mathrm{b}}$ using the rate measured at a reference energy where $\sigma_{\mathrm{I}}$ or $\sigma_{\mathrm{DI}}$ is either zero or known from previous measurements so that the stripping rate can be inferred. For EIDI the background is primarily due to DES, analogous to the situation in the EISI measurement where the background is due to SES. For the double ionization measurement it is possible for the combination of SES plus EISI in sequence to contribute to the background, thereby introducing an energy dependence that cannot be accounted for by measuring the count rate at the fixed reference energy. However, the expected rate for this multiple collision process was estimated based on the measured EISI and SES count rates to be at most about $10^{-6}$ times the rate of DES. Thus, multiple collisions are extremely rare and can be ignored in the analysis.
	
	The background rate $R_{\mathrm{I}}^{\mathrm{b}}$ is proportional to the SES or DES cross section, the number of ions, and the residual gas density of the vacuum. The stripping cross sections depend only on $v_{\mathrm{i}}$, which remains constant throughout the measurement. However, the ion beam decays during the measurement, so there is a slight difference in ion number related to the time delay between the measurement step and the reference step. Additionally, the residual gas density can vary systematically with energy in a way that seems to be related to the probe beam electron current. This can also introduce a systematic distortion of the cross section as a function of energy. Taken all together we expect that $R_{\mathrm{I}}^{\mathrm{b}}(E_{\mathrm{m}}) \neq R_{\mathrm{I}}^{\mathrm{b}}(E_{\mathrm{r}})$.
	
	We correct for both of the above systematic errors in our analysis using the method described in \citet{Hahn:ApJ:2010}. The end result of the correction procedure is to adjust the shape of the cross section by about 3\%. The correction uses the recombination signal at high energies as a proxy for the pressure to correct the reference count rate so that it better reflects the true background rate at measurement. This works because at high energies the recombination signal is dominated by single electron capture (SEC), which depends on residual gas density and ion current in the same way that SES and DES do. The electron capture signal can therefore be used to detect relative differences in pressure and ion current between the reference and measurement steps. The analysis of the recombination data shows that dielectronic recombination (DR) of $\fe$ contributes less than $10\%$ of the total recombination signal above the ionization threshold (Novotn\'{y} et al., in preparation). This measured DR component is subtracted from the total recombination signal in order to use only the SEC signal for the correction. 

	When the reference energy is greater than the ionization threshold the above procedure has to be modified. It is possible to account for the decay of the ion beam using direct measurements of the ion current. The ion beam decays exponentially with a characteristic decay time measured to be $\approx 18$~s. The time between the measurement and reference steps was about $20$~ms. Therefore, the expected difference between the measured reference rate and the actual background rate from stripping at the measurement step due to the decay of the ion beam is only $0.2\%$. The effect of the energy dependent pressure variation was not corrected for scans where the reference energy is above the ionization threshold. Based on the size of the correction applied to the low energy scans, we estimate that not correcting single ionization scans above 950~eV or double ionization scans above 1400~eV introduces a distortion of about 3\% in those energy ranges.  	
	
\section{Uncertainties}\label{sec:err}

	The statistical uncertainty on the cross section is not the same for all energies. For single ionization the statistical uncertainty is about $10\%$ at 400~eV and drops to $1\%$ by 700~eV as the number of counts increases with the increasing ionization cross section. The statistical uncertainty remains at that level until $1400$~eV, where it grows to $\approx 3\%$, because at such high energies the cross section is relatively featureless and we performed only a few low resolution energy scans. The double ionization cross section was measured with a few low resolution scans, and the cross section is much smaller than for single ionization. Consequently, the statistical uncertainties for those results are $\approx 5\%$ on average, with larger uncertainties below 1000~eV where the cross section and corresponding count rates are very small. 
	
	There is a systematic error due to the uncertainty in the stored ion current measurement. Here the ion current was measured non-destructively using a beam profile monitor \citep[BPM;][]{Hochadel:NIMA:1994}. The absolute calibration of this instrument depends on the residual gas pressure and any electronic drifts. The calibration drifts with a timescale that appears to be one to several hours. To correct for the drift of the BPM calibration we performed an initial calibration of the BPM. This calibration was seen to be stable over a three hour period. During this time period we performed a long range energy scan, which covered the energy range of 260~--~950~eV. This scan gives an accurate measurement for the shape of the cross section as a function of energy. For all other data runs we adjusted the ion current calibration in order to produce agreement with this long energy range scan. This normalization is possible because all other quantities were measured much more accurately than the ion current, so we can attribute discontinuities between the energy scans, which overlap in energy range by 50\% and so should give identical results, as being caused by shifts in the ion current calibration. Thus, the details of the cross section can be filled in without introducing distortions to the shape of the curve from the normalization. Finally, we quantified the systematic uncertainty in the magnitude of the cross section by repeating the analysis of the low resolution scans using high and low estimates for the BPM calibration. We estimate the average $1\sigma$ systematic uncertainty from the ion current measurement to be about $12\%$.
	
	Other sources of systematic error were negligible compared to that from the ion current measurement. In the earlier measurements of \citet{Hahn:ApJ:2010} the ion beam was not cooled during measurement and the expansion of the ion beam introduced additional systematic errors. These included possible loss of detection efficiency due to ions far from the center of the beam not hitting the detectors and uncertainties in the toroidal corrections due to the slight differences in the path length through the merging and demerging sections of the interaction region experienced by ions at different transverse locations within the expanded beam. In the present experiment, beam profile measurements using the BPM showed that constant cooling limited the ion beam width to $\approx 1$~mm. Since the width was small compared to the size of the particle detectors, the detection efficiency was essentially unity. Similarly, the systematic uncertainty due to the transverse size of the beam in the merging and demerging sections was also negligible. The uncertainty on the electron density is about 1\% \citep{Kenntner:Thesis}. All uncertainties are summarized in Table~\ref{table:err}.
		
\section{Results for Single Ionization}\label{sec:sres}

\begin{figure}
\centering \includegraphics[width=0.9\textwidth]{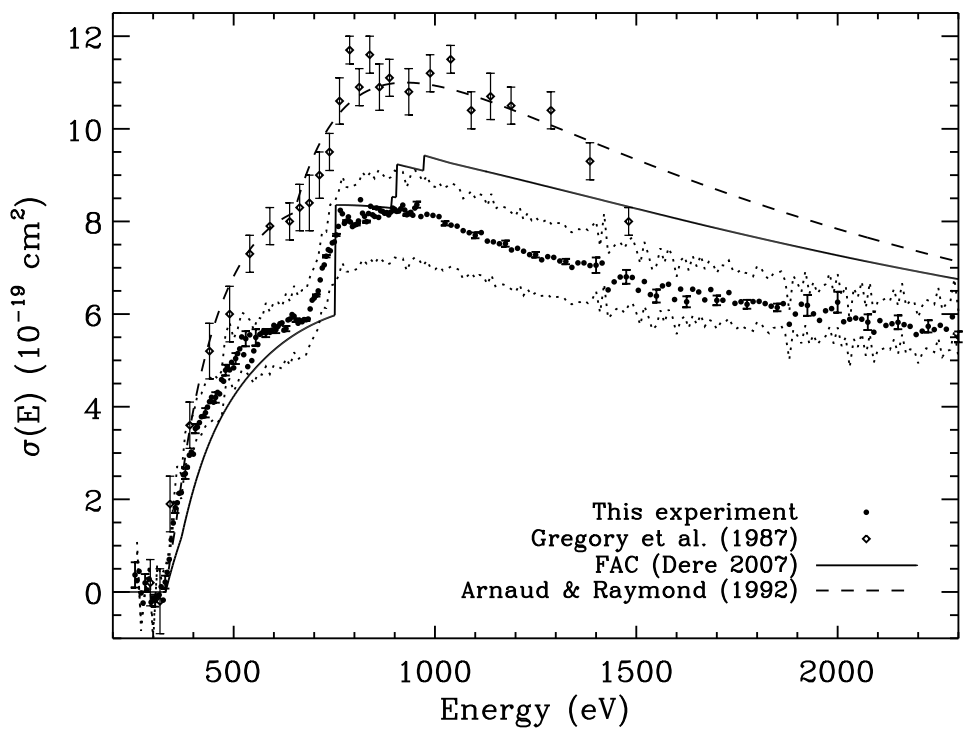}
\caption{\label{fig:fe11single} The EISI cross section for $\fe$ forming $\mathrm{Fe^{12+}}$ is shown here. The filled circles indicate the experimental values and the error bars at selected points illustrate the $1\sigma$ statistical uncertainty. The dotted curves show the $1\sigma$ range of systematic uncertainty from the calibration of the ion current measurement. The diamonds display the experimental results of \citet[][]{Gregory:PRA:1987} and a fit to that data that was used in the CIE calculations of \citet{Arnaud:ApJ:1992} is indicated by a dashed line. The distorted wave calculation of \citet{Dere:AA:2007} using the FAC code is denoted by the solid curve.
}
\end{figure}

	The EISI ionization cross section for $\fe$ forming $\mathrm{Fe^{12+}}$ is shown in Figure~\ref{fig:fe11single}. The dotted curves in the figure indicate the $1\sigma$ systematic uncertainty due to the ion current calibration. The figure also shows the experimental results of \citet{Gregory:PRA:1987}, the fit to that data used in the CIE calculations of \citet{Arnaud:ApJ:1992}, and the Flexible Atomic Code (FAC) distorted wave calculation of \citet{Dere:AA:2007}. 
	
	Figure~\ref{fig:fe11single} shows that the crossed beams measurement of \citet{Gregory:PRA:1987} is about 30\% larger than the current measurement. This is well outside the uncertainties of either experiment, which supports the hypothesis that the earlier measurement had a significant metastable population in the ion beam. Consequently, CSD calculations that relied on those data are likely to be inaccurate \citep[e.g.,][]{Arnaud:ApJ:1992, Mazzotta:AAS:1998}. A recent CSD calculation by \citet{Bryans:ApJ:2009} used the EISI cross section from the FAC calculation of \citet{Dere:AA:2007}. Figure~\ref{fig:fe11single} shows that this calculation is closer to the present experimental results, but significant differences remain. 
		
	In the low energy range from the ionization threshold at approximately 330~eV up to about 690~eV, a comparison of our experimental results and theory indicate that the ionization cross section is dominated by direct ionization. However, the experimental cross section increases faster near threshold than predicted by the calculations. There are two related effects that can account for the faster than expected increase. The cross section may be enhanced due to electron impact excitation of $3s$ electrons to states that relax by autoionization. This ionization mechanism has also been suggested for other P-like ions \citep{Gregory:PRA:1983, Mueller:PRA:1985, Yamada:JPhysJap:1988}. Based on an estimate using the LANL Atomic Code suite \citep{Magee:ASP:1995}, this mechanism is energetically possible for excitation to $n \geq 8$ and is sufficient to account for the increased cross section if the branching ratio for autoionization compared to radiative stabilization is large enough. However, we are unaware of any published data for either EA from $3s$ excitation or the relevant branching ratio. There is also a systematic effect from field ionization that could partially account for the discrepancy. The magnetic fields in TSR cause electric fields in the rest frame of the ions. If a collision excites an electron to a high enough $n$ level it can be field-ionized by the motional electric field generated in the first dipole magnet downstream of the interaction region \citep{Schippers:ApJ:2001}. Here the semi-classical onset for field ionization is $n > 44$. Excitation of a $3p$ electron to $n > 44$ can lead to field ionization that systematically increases the measured cross section. Excitation of a $3s$ electron to $n > 44$ also leads to field ionization in the experiment, but we expect these states primarily to ionize through EA even in the absence of external fields. While the effect of these processes on the cross section is most obvious near the ionization threshold, they also increase the measured ionization cross section for all higher energies.
	
	The relative importance of excitation-autoionization versus field ionization can be estimated from the dependence of the excitation cross section on the $n$ to which the electron is excited. In the Bethe approximation the electron impact excitation cross section for dipole allowed transitions from level $i$ to level $j$ is given by \citep{VanRegemorter:ApJ:1962}
\begin{equation}
\sigma_{\mathrm{exc},ij} = \frac{8\pi}{\sqrt{3}} \frac{I_{\mathrm{H}}}{E \Delta E_{ij}} f_{ij} g \pi a_0^2,
\label{eq:ExcBethe}
\end{equation}
where $E$ is the electron collision energy and $\Delta E_{ij}$ is the threshold energy for the transition, both measured in Rydbergs, $I_{\mathrm{H}}$ is one Rydberg, $f_{ij}$ is the oscillator strength, $g$ is a Gaunt factor, and $a_{0}$ is the Bohr radius. For large $n$ the excitation energy $\Delta E_{ij}$ is approximately constant and equal to the ionization energy. Thus, for a given collision energy $\sigma_{\mathrm{exc},ij} \propto f_{ij}$. For large $n$ the oscillator strength falls off as $1/n^{3}$ \citep{Bethe:Book}. Therefore the ratio of the cross section for excitation to field-ionizing levels relative to autoionizing levels is approximately $\left(2 \sum_{n=44}^{\infty} 1/n^3 \right) / \left(\sum_{n=8}^{\infty} 1/n^3\right) = 0.06$. The factor of two roughly accounts for the fact that excitation of either a $3s$ or a $3p$ electron can lead to field ionization but only excitation of $3s$ can lead to autoionization. This estimate suggests that the systematic contribution to the measured ionization cross section from field ionization is small relative to excitation autoionization, on the order of a few percent only, but detailed calculations taking into account the branching ratios are needed to confirm this. 
			
\begin{figure}
\centering \includegraphics[width=0.9\textwidth]{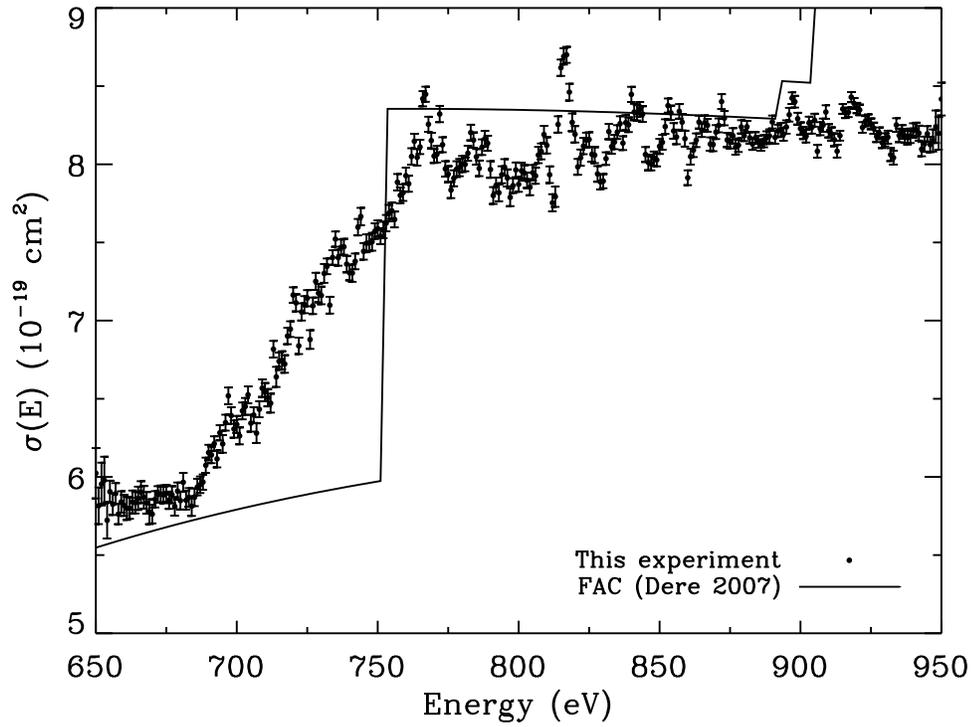}
\caption{\label{fig:fe11finebin} This figure shows the EISI cross section of $\fe$ focussing on the resonances in the range 650 -- 950~eV. The error bars indicate the $1\sigma$ statistical errors, which are small compared to the size of the resonances in this energy range. The distorted wave calculation of \citet{Dere:AA:2007} is also shown.
}
\end{figure}

	From the threshold behavior of the cross section near 700~eV (Figure~\ref{fig:fe11finebin}) we infer that EA of $n=2$ electrons makes up about 25\% of the total cross section at high energies. \citet{Pindzola:PRA:1986} predict EA from $2p \rightarrow 3p$ excitations at $709.3$~eV and from $2p \rightarrow 3d$ excitations at $763.8$~eV. This roughly corresponds with the experimentally observed increase in the cross section beginning at $\approx 690$~eV. The experimentally measured EA from $n=2$ excitations could also be systematically enhanced by field ionization, which inhibits radiative stabilization for excitation to $n > 44$, but this effect should be very small since such states are expected to autoionize even without external fields. The distorted wave calculations of \citet{Dere:AA:2007} show EA of similar magnitude to what was measured experimentally, but starting at somewhat higher energies of $\approx 750$~eV. \citet{Pindzola:PRA:1986} predict additional EA from $2p \rightarrow 4p$, $2p \rightarrow 4d$, and $2s \rightarrow 3d$ excitations to be found in the range 880~--~900~eV and \citet{Dere:AA:2007} also predicts an increase in the cross section due to EA at those energies. However, we do not see a significant increase in the experimental data in that energy range. A possible explanation for this is that the theory has overestimated the branching ratio of the intermediate state for autoionization versus radiative stabilization.
	
	In the energy range of 650 -- 950~eV resonances are observed in the cross section, as shown in more detail in Figure~\ref{fig:fe11finebin}. These are due to the REDA and READI processes which result in single ionization \citep{Laguttuta:PRA:1981,Henry:PRA:1982,Muller:PRL:1988, Muller:PRL:1988b}. 
	
	At collision energies near 1100~eV direct ionization of the $n=2$ electrons becomes possible, but no corresponding feature is observed in the single ionization cross section at this point. This is not surprising since the excited states created by these $L$-shell ionizations are expected to autoionize producing a net double ionization, as is discussed below \citep{Kaastra:AAS:1993}. 
	
	The experimental cross section was used to determine the plasma ionization rate coefficient $\alpha_{\mathrm{I}}(T_{\mathrm{e}})$. This calculation involves multiplying the ionization cross section by the electron-ion relative velocity and integrating over the Maxwellian velocity distribution. Following \citet{Fogle:ApJS:2008} we truncate the integral over energy at $E_0 + 6 k_{\mathrm{B}}T_{\mathrm{e}}$, where $E_0=330.79$~eV is the ionization potential of $\fe$, $T_{\mathrm{e}}$ is the temperature at which $\alpha_{\mathrm{I}}(T_{\mathrm{e}})$ is to be calculated, and $k_{\mathrm{B}}$ is the Boltzmann constant. The cross section was measured only up to $E=2300$~eV. Thus, the rate coefficient for $T_{\mathrm{e}} < 3.8 \times 10^{6}$~K could be calculated without extrapolation. In order to calculate $\alpha_{\mathrm{I}}$ for $T_{\mathrm{e}} > 3.8\times10^{6}$~K we assumed that the falloff of $\sigma_{\mathrm{I}}$ at high energies is the same as that given by the FAC calculation of \citet{Dere:AA:2007} and scale that cross section to our experimental results to extrapolate the measured cross section beyond 2300~eV. The uncertainty on the experimentally-derived rate coefficient is $\pm 12\%$ due primarily to the systematic uncertainty in the ion current calibration. 

\begin{figure}
\centering \includegraphics[width=0.9\textwidth]{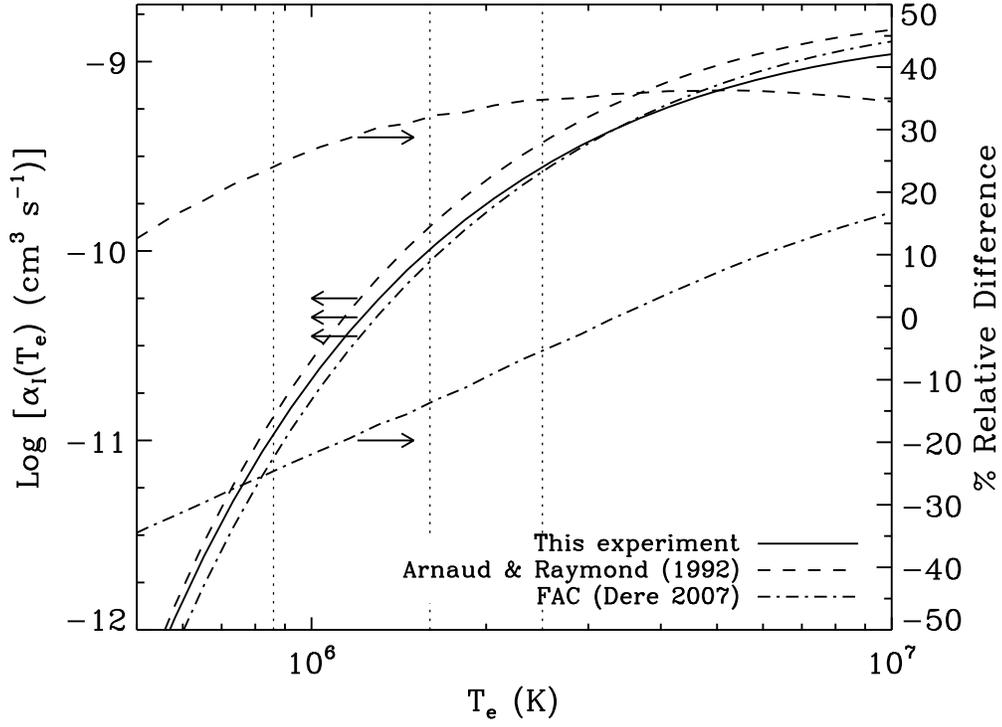}
\caption{\label{fig:ratecoeff} The EISI rate coefficient for $\fe$ forming $\mathrm{Fe^{12+}}$ derived from the experimental data (solid line). Also shown are rate coefficients derived from the earlier experiment of \citet{Gregory:PRA:1987} by Arnaud~\&~Raymond~(1992; dashed line) and from the FAC calculation by Dere~(2007; dash-dotted line). The relative differences between these and the present experimental rate coefficients are also illustrated. Arrows are placed next to the curves to clarify the axis from which the values are to be read off. The vertical dotted lines show the temperature range where the equilibrium ion abundance of $\fe$ is $> 1\%$ with the middle dotted line showing the temperature of maximum ion abundance \citep{Bryans:ApJ:2009}. The systematic error on the experimentally derived rate coefficient is about $\pm12\%$. }
\end{figure}

	Figure~\ref{fig:ratecoeff} compares the plasma rate coefficient from the present measurements to those obtained from the FAC calculation of \citet{Dere:AA:2007} and from the fit to the crossed beams experimental cross section \citep{Gregory:PRA:1987} used in the CIE calculations of \citet{Arnaud:ApJ:1992} and \citet{Mazzotta:AAS:1998}. The ionization equilibrium depends most strongly on the rate coefficients near the peak ion abundance. Recent CSD calculations show that $\fe$ has a greater than 1\% abundance in the temperature range from $8.6 \times 10^{5}$~K to $2.5 \times 10^{6}$~K, with a maximum ion abundance near $1.6 \times 10^{6}$~K \citep{Bryans:ApJ:2009}. The rate coefficient from the \citet{Arnaud:ApJ:1992} fit is about 30\% larger than our results throughout this temperature range. The rate coefficient from the distorted wave calculation is about $15\%$ smaller than the present measurement at the temperature of maximum abundance (Figure~\ref{fig:ratecoeff}). The difference is likely due to the exclusion of EA from the $n=3$ level in the calculation, as well as potential systematic effects of field ionization in the experiment. Since these processes could not be distinguished experimentally, the up to $\sim 25\%$ difference represents an upper bound on the possible effect of EA from $n=3$ levels. It appears that including EA from the $n=3$ level in the calculation of ionization cross sections and rate coefficients is necessary for more accurate EISI rate coefficients.	
	
	In order to produce a fit to the rate coefficient we used the Burgess-Tully type scaling from \citet{Dere:AA:2007} and fit the scaled rate coefficient with a fifth order polynomial. The temperature was scaled as 
\begin{equation}
x = 1 - \frac{\ln 2}{\ln(t + 2)}, 
\label{eq:scaletemp}
\end{equation}
where $t = k_{\mathrm{B}} T_{\mathrm{e}}/E_0$. The rate coefficient was scaled as
\begin{equation}
\rho=t^{1/2}E_0^{3/2}\alpha_{\mathrm{I}}(T_{\mathrm{e}})/E_{1}(1/t),
\label{eq:scalerate}
\end{equation}
where $E_{1}(1/t)$ is the first exponential integral. The scaled temperature $x$ and rate coefficient $\rho$ can be inverted to reproduce $T_{\mathrm{e}}$ and $\alpha_{\mathrm{I}}$ using 
\begin{equation}
T_{\mathrm{e}} = \frac{E_0}{k_{\mathrm{B}}}\left[\exp\left(\frac{\ln 2}{1-x} \right) - 2 \right] 
\label{eq:invscaletemp}
\end{equation}
and
\begin{equation}
\alpha_{\mathrm{I}}(T_{\mathrm{e}}) = t^{-1/2}E_0^{-3/2}E_{1}(1/t)\rho.
\label{eq:invscalerate}
\end{equation}
The coefficients for a fifth order polynomial fit to $\rho(x)$ are given in Table~\ref{table:coeff}. The fit reproduces the experimental $\alpha_{\mathrm{I}}(T_{\mathrm{e}})$ to 1\% or better accuracy over the temperature range $T_{\mathrm{e}} = 2 \times 10^{5} - 1 \times 10^{8}$~K.
	
\section{Results for Double Ionization}\label{sec:dres}
	
\begin{figure}
\centering \includegraphics[width=0.9\textwidth]{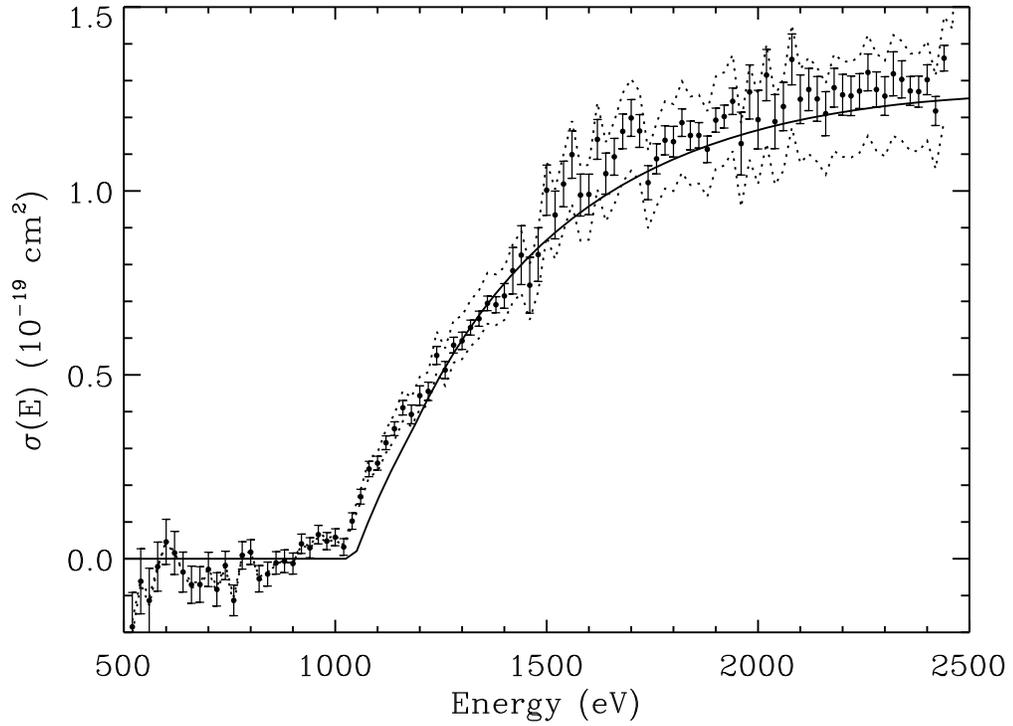}
\caption{\label{fig:fe11double} The EIDI cross section for $\fe$ forming $\mathrm{Fe^{13+}}$ is shown here. The $1\sigma$ statistical uncertainties are indicated by the error bars. The dotted curves illustrate the systematic $1\sigma$ uncertainty from the ion current measurement. Also shown is a calculation using the GIPPER code \citep{Magee:ASP:1995} for the double ionization cross section due solely to single ionization from an $L$-shell electron followed by autoionization. 
}
\end{figure}

	The double ionization cross section in the energy range 500 -- 2500~eV is shown in Figure~\ref{fig:fe11double}. The points show the experimental values. The dotted lines indicate the $1\sigma$ systematic uncertainty from the BPM calibration. The threshold for direct double ionization is 698.83~eV \citep{NIST:2008}, but the measured cross section is not significantly different from zero below the threshold for single ionization of $n=2$ electrons near 1000~eV. 
	
	The solid line in Figure~\ref{fig:fe11double} shows the cross section for single ionization of an $n=2$ electron calculated in the distorted wave approximation using the GIPPER code of the LANL Atomic Physics Code suite \citep{Magee:ASP:1995} scaled to reflect the probability of autoionization relative to radiative stabilization. The scaling factors used were 0.99 for ionization of a $2p$ electron and 0.92 for ionization of a $2s$ electron \citep{Kaastra:AAS:1993}. Thus, the solid line represents the cross section for double ionization proceeding through EISI of an $L$-shell electron followed by autoionization when the ``hole'' is filled in. The EIDI cross section is clearly dominated by $L$-shell single ionization followed by autoionization, as seen by the good agreement between the experimental cross section and the calculated cross section for this process. Such indirect processes are known to become increasingly important relative to direct ionization as the charge state of the ion increases \citep{Muller:PRL:1980, Muller:JphysB:1985, Stenke:JphysB:1999}. 
		
\section{Summary}\label{sec:sum}	
	
	We have measured the EISI and EIDI cross sections of $\fe$. In the single ionization case we found that the cross section was $\approx 30\%$ smaller than that measured in a single-pass, crossed beams experiment \citep{Gregory:PRA:1987}. The difference is believed to be due to the presence of metastable ions in the single-pass experiment and the absence in the current experiment. The single ionization cross section is in agreement with the theoretical cross section of \citet{Dere:AA:2007} to within about $\pm$20\%. The remaining differences are probably due to differences in EA between the calculation and the experiment. At low energies the calculation leaves out EA from excitation of $3s$ electrons. At higher energies the calculation seems to overestimate the EA contribution. Our results show that the Maxwellian EISI rate coefficient from calculations of \citet{Dere:AA:2007} may be underestimated by as much as $\sim25\%$ over the temperatures at which $\fe$ is abundant, partially due to the need to include EA from the $n=3$ level in the calculations. The experimental results for double ionization show that the cross section is dominated by single ionization of an $L$-shell electron which is followed by autoionization. This is consistent with previous results for lower charge state ions, which showed these processes tend to become more important as the ion charge state increases.

\begin{acknowledgments}
	We appreciate the efficient support by the MPIK accelerator and TSR groups during the beamtime. We also thank K. Dere for stimulating discussions. This work was supported in part by the NASA Astronomy and Physics Research and Analysis program and the NASA Solar Heliospheric Physics program. We also acknowledge financial support by the Max-Planck Society, Germany.
\end{acknowledgments}

\appendix

\section{Correction for EII at the Reference Energy}\label{ap:ref}
	
	Equation~(\ref{eq:sigmacalc}) includes a correction term that accounts for a non-zero EII cross section at the reference energy. This term includes a relativistic factor that has been omitted in previous work. This appendix briefly explains the origin of this term. 
		
		The ionization count rate at the measurement energy $E_{\mathrm{m}}$ is given by 
\begin{equation}
R^{\mathrm{m}}_{\mathrm{I}}(E_{\mathrm{m}})=\avg{\sigma_{\mathrm{I}}v_{\mathrm{rel}}}(E_{\mathrm{m}})n_{\mathrm{e}}^{\mathrm{m}}N_{\mathrm{i}}^{\mathrm{m}}L[1-\beta_{\mathrm{i}}\beta_{\mathrm{e}}(E_{\mathrm{m}})] + R_{\mathrm{ES}}. 
\label{eq:meascount}
\end{equation}
Here $R_{\mathrm{ES}}$ represents the background count rate due to electron stripping off the rest gas. The other symbols are defined in the discussion following equation~(\ref{eq:sigmacalc}). For simplicity we ignore pressure fluctuations and the decay of the ion beam and assume that $R_{\mathrm{ES}}$ and $N_{\mathrm{i}}$ are the same at the measurement and reference energies. Systematic errors arising from this assumption can be corrected for separately using the method described in \citet{Hahn:ApJ:2010}. With these simplifications the background count rate measured at the reference energy is given by
\begin{equation}
R^{\mathrm{b}}_{\mathrm{I}}(E_{\mathrm{m}})=\avg{\sigma_{\mathrm{I}}v_{\mathrm{rel}}}(E_{\mathrm{r}})n_{\mathrm{e}}^{\mathrm{r}}N_{\mathrm{i}}^{\mathrm{m}}L[1-\beta_{\mathrm{i}}\beta_{\mathrm{e}}(E_{\mathrm{r}})] + R_{\mathrm{ES}}.
\label{eq:refcount}
\end{equation}
Subtracting equation~(\ref{eq:refcount}) from equation~(\ref{eq:meascount}) and solving for $\avg{\sigma_{\mathrm{I}}v_{\mathrm{rel}}}(E_{\mathrm{m}})$ gives
\begin{equation}
\avg{\sigma_{\mathrm{I}}v_{\mathrm{rel}}}(E_{\mathrm{m}}) = \frac{ R_{\mathrm{I}}^{\mathrm{m}}(E_{\mathrm{m}}) - R_{\mathrm{I}}^{\mathrm{b}}(E_{\mathrm{m}})}{[1 - \beta_{\mathrm{i}}\beta{\mathrm{e}}(E_{\mathrm{m}})] n_{\mathrm{e}}^{\mathrm{m}}N_{\mathrm{i}}^{\mathrm{m}}L} + \avg{\sigma_{\mathrm{I}}v_{\mathrm{rel}}}(E_{\mathrm{r}})\frac{n_{\mathrm{e}}^{\mathrm{r}}}{n_{\mathrm{e}}^{\mathrm{m}}}
\frac{[1-\beta_{\mathrm{i}}\beta_{\mathrm{e}}(E_{\mathrm{r}})]}{[1-\beta_{\mathrm{i}}\beta_{\mathrm{e}}(E_{\mathrm{m}})]}.		
\label{eq:ionratecoeff}
\end{equation}
We recover equation~(\ref{eq:sigmacalc}) after dividing by the relative velocity. In terms of $v/c$, the velocities in the present experiment were small and $[1-\beta_{\mathrm{i}}\beta_{\mathrm{e}}(E)] \approx 0.99$. Hence this additional relativistic correction has a nearly negligible effect.

\section{List of Abbreviations}\label{ap:abb}

\begin{tabbing}
BPM \qquad\= Beam Profile Monitor \\
CEM \> Channel Electron Multiplier \\
CIE \> Collisional Ionization Equilibrium \\
CSD \> Charge State Distribution \\
DES \> Double Electron Stripping \\
DR \> Dielectronic Recombination \\
DW \> Distorted Wave \\
EA \> Excitation Autoionization \\
EIDI \> Electron Impact Double Ionization \\
EII \> Electron Impact Ionization \\
EISI \> Electron Impact Single Ionization \\
REDA \> Resonanat Excitation Double Autoionization \\
READI \> Resonant Excitation Auto-Double-Ionization \\
SEC \> Single Electron Capture  \\
SES \> Single Electron Stripping \\
\end{tabbing}


\clearpage


\begin{center}
\begin{longtable}{lcc}
\caption{Sources of Uncertainty. \label{table:err}}\\
\hline \hline
 & Source & Estimated $1\sigma$ Uncertainty \\
 \hline
 \endfirsthead
 \hline \hline
 & Source & Estimated $1\sigma$ Uncertainty \\
 \hline
 \endhead
 \hline \hline
 \endlastfoot
&Ion current measurement & 12\% \\
&Electron density & 1\% \\
&Detection efficiency\tablenotemark{1} & $< 0.1\%$ \\
&Interaction length spread\tablenotemark{1} & $< 0.1\%$ \\
&Counting statistics & 1\% -- 10\% \\
&Background correction\tablenotemark{2} & 0 -- 3\% \\
\hline
&Quadrature sum & 12\% -- 16\% \\
\footnotetext[1]{Although negligible, these uncertainties are listed here for comparison with the similar experiment of \citet{Hahn:ApJ:2010} where the expansion of the ion beam introduced small additional uncertainties. In the present experiment continuous cooling of the ion beam inhibited beam expansion making these measurements more precise.}
\footnotetext[2]{The 3\% uncertainty applies to data for $E > 950$~eV for single ionization and $E > 1400$~eV for double ionization where the background correction could not be applied.}
\end{longtable}
\end{center}

\clearpage

\begin{center}
\begin{longtable}{ll}
\caption{Fifth-order Polynomial Fitting Parameters to Reproduce the Scaled Ionization Rate Coefficient $\rho(x)$ (see equations \ref{eq:scaletemp} and \ref{eq:scalerate}). \label{table:coeff}} \\
\hline \hline 
$i$ & $a_{i}$  \\
\hline
\endfirsthead
\hline \hline 
$i$ & $a_{i}$  \\
\hline
\endhead
\hline\hline
$\displaystyle \rho = \sum_{i=0}^{i=5}{a_{i}x^{i}}$
\endlastfoot
0 & \phs$19.4438$ \\
1 & $-76.2632$ \\
2 & \phs$376.369$ \\
3 & $-819.161$ \\
4 & \phs$827.141$ \\
5 & $-322.187$ \\
\end{longtable}
\end{center}

\bibliography{TSR_bib}

\end{document}